\documentclass[reprint,aps,pra]{revtex4-1}

\usepackage{graphicx,epsfig}
\usepackage[colorlinks=true,allcolors=blue]{hyperref}
\usepackage{amssymb}

\newcommand{\CMA}{$\mathrm{Co_2MnAl}$\ }
\newcommand{\FMA}{$\mathrm{Fe_2MnAl}$\ }
\newcommand{\CMAFMA}{$\mathrm{[CMA_n/FMA_n]_q}$\ }
\newcommand{\FMACMA}{$\mathrm{[FMA_n/CMA_n]_q}$\ }

\newcommand{\mats}{Materials Department, University of California, Santa Barbara, California, 93106, USA}
\newcommand{\ece}{Dept. of Electrical and Computer Engineering, University of California, Santa Barbara, California, 93106, USA}
\newcommand{\lorraine}{Institut Jean-Lamour, UMR 7198, CNRS, Universit\'e Lorraine, Nancy, France}
\newcommand{\soleil}{Synchrotron SOLEIL, L'Orme des Merisiers, Saint-Aubin, BP 48, F-91192 Gif sur Yvette, France}

\begin{document}

\title{Epitaxial Heusler Superlattice Co$_2$MnAl/Fe$_2$MnAl with Perpendicular Magnetic Anisotropy and Termination-Dependent Half-Metallicity} 




\author{Tobias L. Brown-Heft}
\affiliation{\mats}

\author{Anthony P. McFadden}
\affiliation{\ece}

\author{John A. Logan}
\affiliation{\mats}

\author{Charles Guillemard}
\affiliation{\lorraine}
\affiliation{\soleil}

\author{Patrick Le F\`evre}
\affiliation{\soleil}

\author{Fran\c cois Bertran}
\affiliation{\soleil}

\author{St\'ephane Andrieu}
\affiliation{\lorraine}

\author{Chris J. Palmstr\o m}
\email[]{cpalmstrom@ece.ucsb.edu}
\affiliation{\mats}
\affiliation{\ece}


\begin{abstract}
Single-crystal Heusler atomic-scale superlattices that have been predicted to exhibit perpendicular magnetic anisotropy and half-metallicity have been successfully grown by molecular beam epitaxy. Superlattices consisting of full-Heusler \CMA and \FMA with one to three unit cell periodicity were grown on GaAs~(001), MgO~(001), and Cr~(001)/MgO~(001). Electron energy loss spectroscopy maps confirmed clearly segregated epitaxial Heusler layers with high cobalt or high iron concentrations for samples grown near room temperature on GaAs~(001). Superlattice structures grown with an excess of aluminum had significantly lower thin film shape anisotropy and resulted in an out-of-plane spin reorientation transition at temperatures below 200~K for samples grown on GaAs~(001). Synchrotron-based spin resolved photoemission spectroscopy found that the superlattice structure improves the Fermi level spin polarization near the $X$ point in the bulk Brillouin zone. Stoichiometric \CMA terminated superlattice grown on MgO~(001) had a spin polarization of 95\%, while a pure \CMA film had a spin polarization of only 65\%.
\end{abstract}


\maketitle 

\section{Introduction}
Spintronic devices require a source of spin-polarized current, and ferromagnetic metals are commonly used for this purpose due to their imbalance of spin up and spin down electron density of states near the Fermi level \cite{Graf2011}. Two physical phenomena useful for the improvement of ferromagnetic electrodes used in magnetic tunnel junctions are perpendicular magnetic anisotropy (PMA) \cite{Ikeda2010} and half-metallicity \cite{Jourdan2014,Andrieu2016}. Extensive research has been conducted to realize these features independently, for example in CoFeB/MgO \cite{Ikeda2010}, $\mathrm{Co_2MnSi}$ \cite{Jourdan2014}, and $\mathrm{Co_2MnSi/MgO}$ \cite{Andrieu2016}. Others have combined separate material systems into hybrid electrodes where a thin half-metal is magnetically pinned in the out-of-plane direction by an adjacent layer with strong PMA \cite{Hiratsuka2010}. The compensated ferrimagnet $\mathrm{Mn_2Ru_xGa}$ can be integrated into perpendicular magnetic tunnel junctions and is predicted to be half-metallic under specific conditions \cite{Borisov2017,Zic2016}. However, thus far, a single material exhibiting both PMA and half-metallicity has yet to be experimentally confirmed. In this work, we present a promising Heusler atomic superlattice that exhibits both PMA and half-metallicity, albeit for separate samples with different growth conditions.

Half-metals are ferromagnets that possess a Slater-Pauling gap in one of the spin directions and a Fermi level position that lies within that gap. Consequently, they behave like a metal for one spin channel and a semiconductor for the other, resulting in 100\% spin-polarized conduction electrons. Many half-metals have been predicted to exist within the cobalt-based full-Heusler family of materials. Full-Heuslers have molecular formula $\mathrm{X_2YZ}$, where X and Y are typically \textit{d}- or \textit{f}-block elements and Z is typically an \textit{sp} element. In the ideal $\mathrm{L2_1}$ crystal structure, the Y and Z atoms form a rocksalt lattice that is filled with X atoms in each of the eight tetrahedral sites, resulting in $Fm\bar{3}m$ space group symmetry. Full-Heusler compounds also commonly crystalize in the B2 (CsCl) structure that represents disorder between the Y and Z atomic sites, which changes to $Pm\bar{3}m$ space group symmetry. In both cases, full-Heusler compounds possess cubic symmetry that gives rise to cubic magnetocrystalline anisotropy that, on its own, cannot overcome thin film magnetic shape anisotropy to yield PMA.

Recently, it was predicted that atomic superlattices of certain pairs of Heusler materials could be perpendicularly magnetized half-metals \cite{Azadani2016}. Heusler superlattices are distinct from other magnetic multilayers because they maintain the same crystal structure and, in many cases, several of the same atomic species in both constituent layers. The uniaxial anisotropy in the growth direction arises from changes in electronic structure between layers, and from lattice distortions produced by variations in lattice constant between parent bulk crystals \cite{Azadani2016}. Equivalently, symmetry breaking due to the layer structure results in tetragonal space group symmetry, which gives rise to tetragonal magnetic anisotropy with the unique axis aligned out-of-plane. In addition, the mixing of electronic states across sublayers is calculated to have a Fermi level tuning effect. Two Heusler compounds that are not half-metallic may combine into a superlattice with the Fermi level within the Slater-Pauling gap, forming a half-metal \cite{Azadani2016}.

\begin{figure}[h]
\includegraphics[width=.9\linewidth]{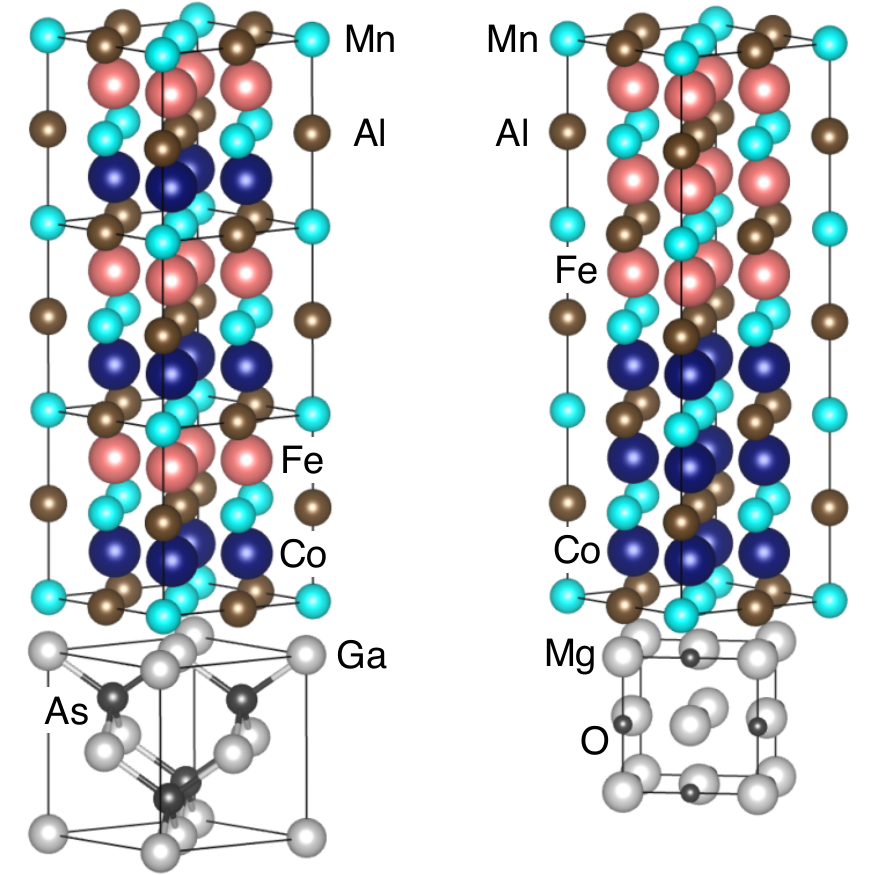}
\caption{\label{crystal}Schematics of crystal structure and epitaxial relationship for (left) $\mathrm{[CMA_{0.5}/FMA_{0.5}]_3/GaAs~(001)}$ and (right) $\mathrm{[CMA_{1.5}/FMA_{1.5}]_1/MgO~(001)}$ viewed along the Heusler [110] direction.}
\end{figure}

The superlattice composed of \CMA (CMA) and \FMA (FMA) layered along the [001] direction is predicted to exhibit both PMA and half-metallicity for specific superlattice periodicities. We adopt the convention of Azadani \textit{et al.} \cite{Azadani2016} and grow superlattices with nominal layering of n~=~0.5 and n~=~1.5, where n is the thickness of each CMA or FMA sublayer in fractions of a Heusler unit cell. These are stacked to produce \CMAFMA films, where q is the total number of bilayers in the superlattice, as shown in Fig.~\ref{crystal}. Defined in this way, the reduced space group symmetry of the superlattice is $P4/nmm$ \cite{Stokes2005}. However, in this work, all Heusler lattice parameters and Bragg reflections are given in terms of the $\mathrm{L2_1}$ structure.

\section{Experiment}
The \CMAFMA films were deposited on GaAs~(001), MgO~(001), and chromium-buffered Cr~(001)/MgO~(001) via molecular beam epitaxy in a modified Veeco Gen~II growth chamber with base pressure $\mathrm{<5\times10^{-11}~Torr}$. For growth on GaAs, epi-ready GaAs~(001) wafers were prepared by thermal desorption of the surface oxide under $\mathrm{As_4}$ overpressure in a VG V80H growth chamber, after which a GaAs buffer was grown. After cooling, a sacrificial arsenic capping layer was deposited \textit{in-situ}. The wafer was then loaded out of ultra-high vacuum (UHV) and stored in inert atmosphere. Before growing a Heusler film, a cleaved section of the As/GaAs~(001) wafer was loaded back into UHV where the arsenic cap was thermally desorbed, resulting in a (2x4)/c(2x8) reconstruction in reflection high energy electron diffraction (RHEED). For growth on MgO, MgO (001) substrates were annealed at $\mathrm{800^{\circ}C}$ for 12 hours in an oxygen ambient furnace to reduce root mean square (RMS) surface roughness to ~2~\AA\ \cite{Zama2001}. The MgO substrates were then annealed in UHV at $\mathrm{600^{\circ}C}$ for 30~min, followed by deposition of a 10~nm thick MgO buffer layer at $\mathrm{530^{\circ}C}$ substrate temperature by e-beam evaporation of stoichiometric source material to bury any remaining surface contamination. For MgO with a chromium buffer layer, a 25~nm thick chromium layer was then deposited from a standard effusion cell onto the prepared MgO (001) substrate held at room temperature. The Cr/MgO (001) was subsequently annealed at $\mathrm{500^{\circ}C}$ for 45~min until the surface became smooth, as indicated by streaky RHEED patterns. \textit{In-situ} scanning tunneling microscopy of the annealed chromium surface showed atomic steps and RMS roughness of 1.2~\AA, which is a favorable starting surface for growth of magnetic tunnel junction layers.

The Heusler films were grown by co-evaporation of elemental source material from standard effusion cells. Superlattices were grown by setting atomic fluxes such that \mbox{$\Phi_{Co}=\Phi_{Fe}=2\Phi_{Mn}=2\Phi_{Al}$}. In addition, some samples were grown with an increased aluminum flux up to 50\% excess, while keeping other fluxes constant. This allowed for constant co-deposition of the MnAl rocksalt sublattice, while shutters were used to select either cobalt or iron to grow CMA or FMA, respectively. Fluxes were calibrated before each growth using a beam flux gauge mounted to the sample manipulator. The beam equivalent pressure of each effusion cell was calibrated to its true atomic flux calculated from measurements of total elemental atomic layer deposition using Rutherford backscattering spectrometry (RBS) on MgO calibration samples. Superlattice \CMAFMA films with periodicity n~=~0.5~and~1.5 were grown with q~=~34~and~12 full periods, respectively, which gave a film slightly over 20~nm thick in each case. Growth temperatures depended on the substrate chosen, and will be discussed in the following section. 

During Heusler growth, surface crystal quality was monitored by \textit{in-situ} RHEED. After growth, samples were capped with 10~nm $\mathrm{AlO_x}$ deposited by \textit{in-situ} e-beam evaporation of $\mathrm{Al_2O_3}$ source material to prevent film oxidation, and loaded out of UHV for \textit{ex-situ} characterization. Film morphology was measured with atomic force microscopy (AFM) in tapping mode. Initial crystal quality was measured by Cu $\mathrm{K_{\alpha 1}}$ X-ray diffraction (XRD) open detector rocking curves, while lattice parameters were extracted from XRD reciprocal space maps (RSM) collected with a CCD line detector. Magnetic hysteresis loops were collected using a Quantum Design MPMS~XL SQUID. Film thicknesses were measured by X-ray reflectometry (XRR). Sample areas were determined photographically.

\section{Results and Discussion}
\subsection{Structural Quality}
\begin{figure}[h]
\includegraphics[width=.9\linewidth]{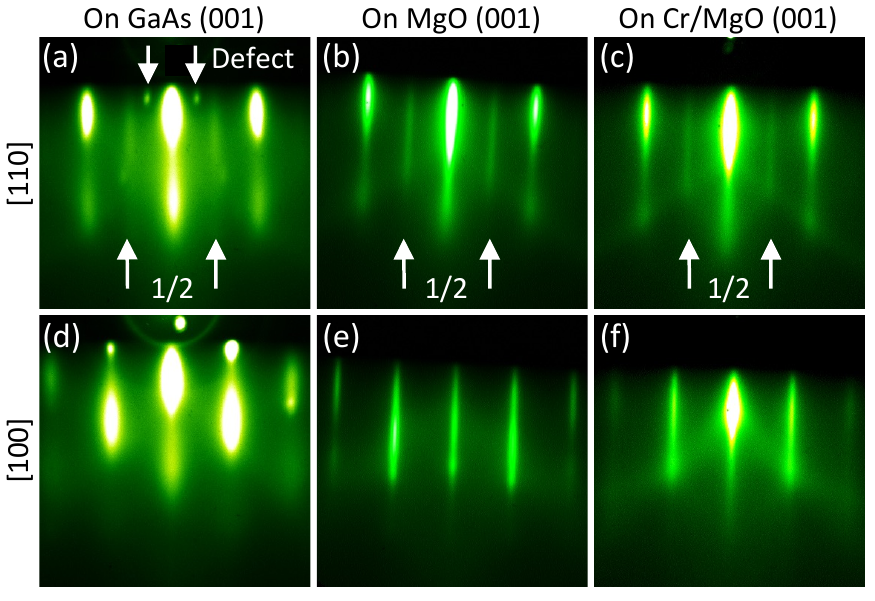}
\caption{\label{RHEED}RHEED images of 20~nm thick \CMAFMA films grown on indicated substrates along Heusler (a-c) [110] and (d-f) [100] directions. Half-order streaks along [110] indicate an $\mathrm{L2_1}$-like surface unit cell. Faint spots visible in (a), which vanish after annealing to $\mathrm{300^{\circ}C}$, are attributed to crystal twinning at low growth temperatures.}
\end{figure}

\begin{figure}[h]
\includegraphics[width=.9\linewidth]{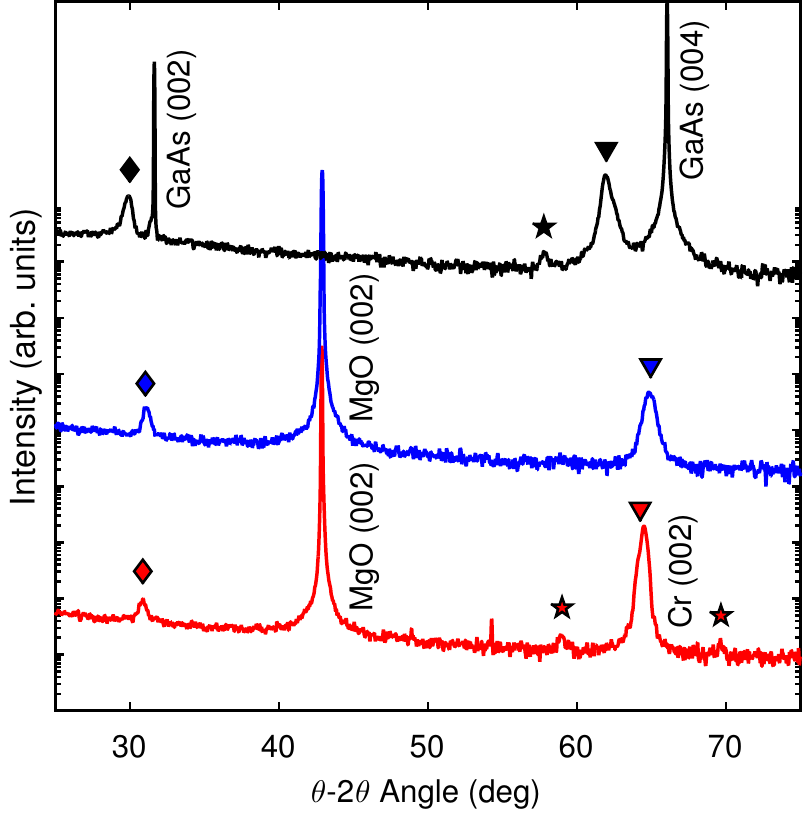}
\caption{\label{XRD}XRD on-axis rocking curves for $\mathrm{[CMA_{1.5}/FMA_{1.5}]_{12}}$ films grown on (black line) GaAs~(001), (blue line) MgO~(001), and (red line) Cr/MgO~(001). Atomic ordering that is at least B2 is confirmed by the presence of a Heusler (002) peak ($\blacklozenge$) along with the (004) peak ($\blacktriangledown$). The superlattice satellite peak ($\bigstar$) corresponds to a periodicity of 24.6~\AA\ for the film grown on GaAs~(001), and 20.7~\AA\ for the film grown on Cr/MgO~(001). Only one satellite peak was distinguishable for films grown on GaAs~(001) due to overlap with the substrate (004) peak. Satellite peaks were not observed for films grown on MgO~(001) due to film roughness and/or diffusion effects.}
\end{figure}

For growths on GaAs (001), \CMAFMA films were grown from $\mathrm{150^{\circ}C}$ to $\mathrm{300^{\circ}C}$ substrate temperature, resulting in films with a (002) Bragg reflection in XRD indicating at least partial B2 ordering as shown in Fig.~\ref{XRD}. Fast diffusion of adatoms along arsenic dimer rows produced corrugations visible in AFM along GaAs $\mathrm{[1\bar{1}0]}$, which resulted in root mean square (RMS) surface roughness of 8.3~\AA\ for the final $\mathrm{AlO_x}$-capped Heusler films. The roughness is also apparent in RHEED images, indicated by spottiness along the diffraction streaks as shown in Fig.~\ref{RHEED}(a) and (d). For growths on MgO (001), \CMAFMA films with a (002) reflection present were obtained both for samples grown at $\mathrm{300^{\circ}C}$, and for those grown at room temperature and subsequently annealed at $\mathrm{300^{\circ}C}$ for 15~min. Islands 40~nm wide and 1 to 4~nm tall visible in AFM resulted from surface energy mismatch and 2.9\% tensile film strain, giving an RMS roughness of 6.0~\AA. These islands were also present for A2 (bcc solid solution) films lacking a (002) Bragg reflection, grown at room temperature with no subsequent anneal, suggesting the island morphology was not caused by dewetting at high temperatures. RHEED images showing a c(2x2) reconstruction with prominent half-order streaks along [110] indicated high quality Heusler growth and suggested an $\mathrm{L2_1}$-like surface unit cell \cite{Dong1999}. Finally, \CMAFMA grown on Cr/MgO~(001) at $\mathrm{250^{\circ}C}$ had a (002) Bragg reflection and exceptionally smooth surface morphology with 2.4~\AA~RMS roughness. Bright half-order streaks and Kikuchi lines in RHEED images confirmed smooth surfaces and high crystal quality suitable for fabrication of devices such as magnetic tunnel junctions. These results are summarized in Table~\ref{table}.

XRD reciprocal space maps of the \CMAFMA (224) reflections were collected along with a nearby substrate reflection. Using these off-axis peaks, the in-plane and out-of-plane lattice parameters of the superlattice were calculated. The \CMAFMA films were partially to fully strained to the substrates, with the degree of relaxation increasing slightly with higher growth and annealing temperatures. This resulted in tetragonal distortion c/a~=~1.02 to 1.06 for films deposited on GaAs~(001), c/a~=~0.96 to 0.99 for films deposited on MgO~(001), and c/a~=~1.00 for films deposited on lattice matched Cr/MgO~(001). The \CMAFMA relaxed cubic lattice parameter $\mathrm{a_0=5.79}$~\AA\ was extracted from a linear fit of a~vs.~c.

\begin{figure}[t]
\includegraphics[width=.49\linewidth]{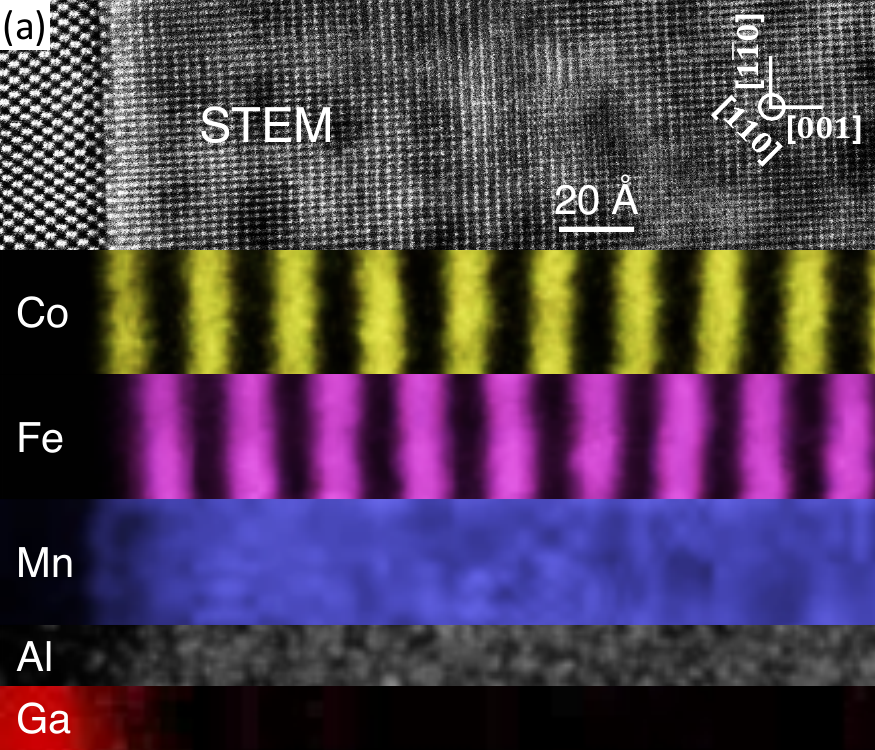}
\includegraphics[width=.49\linewidth]{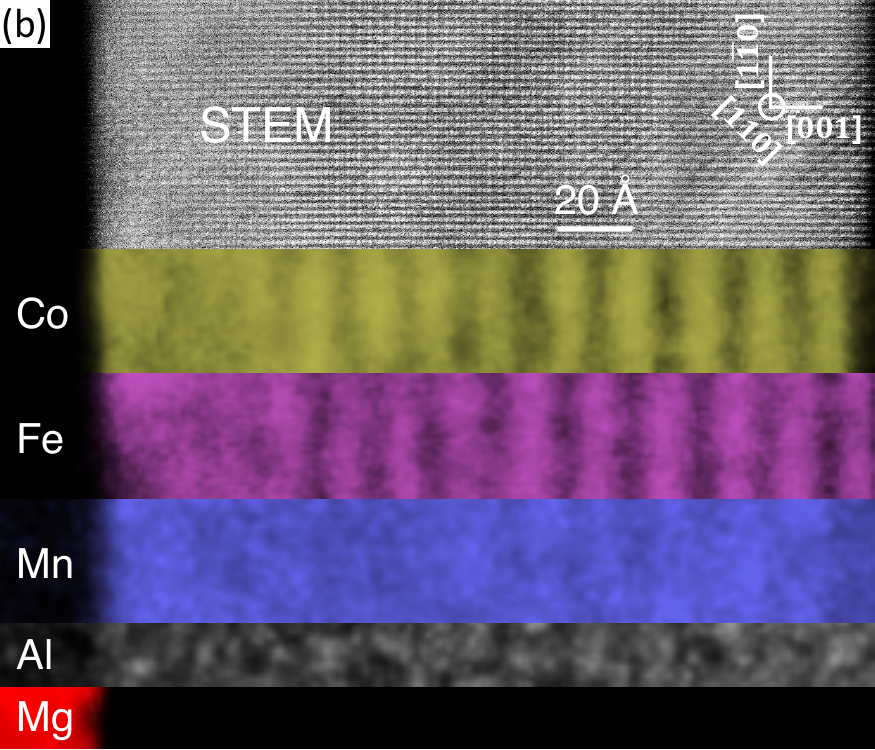}
\caption{\label{EELS}(Top row) Cross-sectional HAADF-STEM and (bottom five rows) EELS maps along the [110] direction of $\mathrm{[CMA_{1.5}/FMA_{1.5}]_{12}}$ superlattices grown on (a) GaAs~(001) at $\mathrm{150^{\circ}C}$ substrate temperature, and (b) MgO~(001) at $\mathrm{300^{\circ}C}$ substrate temperature. A weak superlattice satellite peak was observed in XRD rocking curves for (a) but not for (b).}
\end{figure}

Based on diffraction structure factor calculations, the presence of a Heusler (111) Bragg reflection indicates at least partial $\mathrm{L2_1}$ ordering. A (111) reflection was observed in XRD RSMs for pure \FMA films but was not observed for any \CMA or superlattice films. Additionally, cross-sectional high angle annular dark field scanning transmission electron microscope (HAADF-STEM) images shown in Fig.~\ref{EELS}(a) indicate B2 ordering for a $\mathrm{[CMA_{1.5}/FMA_{1.5}]_{12}/GaAs~(001)}$ film grown at $\mathrm{150^{\circ}C}$ substrate temperature. This is apparent from the lack of a characteristic brickwork pattern expected from the alternating manganese and aluminum atomic columns when viewed along the [110] direction in the $\mathrm{L2_1}$ structure \cite{McFadden2017}, as shown in Fig.~\ref{crystal}. Nevertheless, diffuse half-order streaks observed in RHEED along the Heusler [110] direction during and after growth suggest the surface unit cell is at least partially $\mathrm{L2_1}$-like. STEM electron energy loss spectroscopy (STEM-EELS) maps of the same region reveal that the superlattice structure is intact, cobalt and iron interdiffusion is low, and the thin \CMA -- GaAs interface layer is gallium and cobalt rich, which could indicate an epitaxial CoGa B2 interfacial layer \cite{Palmstrom1989}. 

This analysis was then repeated for \CMAFMA grown at $\mathrm{300^{\circ}C}$ on MgO~(001). Spottiness along the diffraction streaks observed in RHEED during superlattice nucleation suggested an island growth mode and possibly the presence of microtwins, which can form during island coalescence due to slight misorientations between neighboring islands \cite{Yamada1987}. A (111) Bragg reflection was not observed in XRD RSMs, suggesting that the films are B2 ordered. However, STEM shown in Fig.~\ref{EELS}(b) reveals some regions with the characteristic $\mathrm{L2_1}$ brickwork pattern, but the pattern is not uniform across the image, suggesting mixed B2/$\mathrm{L2_1}$ order. Additionally, a disordered region is visible within 2--3~nm of the interface with MgO. Crystallites in this region had small, random rotational mismatches, which were likely caused by the large 2.9\% tensile lattice strain. This disorder is best described as mosaic rather than polycrystalline nucleation. STEM-EELS measured complete sublayer intermixing in this interfacial region. The superlattice structure became visible further from the MgO interface, but significant apparent sublayer intermixing remained. There are two explanations for this behavior. First, diffusion of cobalt and iron within the Heusler matrix during growth could cause sublayers to mix. Layers near the MgO interface were exposed to $\mathrm{300^{\circ}C}$ for one hour longer than those near the surface, which could result in the observed mixing gradient. Second, the island growth morphology indicated by RHEED during Heulser growth initiation may produce height variations greater than the thickness of individual sublayers near the MgO interface. As the film becomes thicker and smoother, the roughness may drop below the sublayer thickness, allowing a superlattice structure to be observed in STEM-EELS. If the average island size is much smaller than the TEM sample thickness, this growth mode would be imaged in EELS as a mixing gradient between the bottom and top surfaces of the superlattice.

The superlattice structures were further analyzed using superlattice satellite peaks observed in XRD. The satellite peaks are expected to be weak because the X-ray scattering form factors of cobalt and iron are quite similar. The satellite peak for the film grown on GaAs~(001) shown Fig.~\ref{XRD} corresponds to a periodicity of 24.6~\AA, which matches well with the periodicity measured by EELS in Fig.~\ref{EELS}(a) for the same sample. The periodicity was larger than the expected $2 n a_0=17.4$~\AA\ primarily because the sample was grown with an aluminum excess of $\mathrm{x=33\%}$ (see section III.B.). Satellite peaks were observed for all superlattice samples with n = 1.5 grown on GaAs~(001) at $\mathrm{200^{\circ}C}$ or below. The absence of a satellite peak for films with higher growth and annealing temperatures suggests that sublayer interdiffusion degrades the superlattice structure. A satellite peak was also observed for a film grown on Cr/MgO~(001) at $\mathrm{250^{\circ}C}$, suggesting that the superlattice structure survives up to slightly higher growth temperatures than for films grown on GaAs~(001). However, films grown or annealed above $\mathrm{300^{\circ}C}$ on Cr/MgO~(001) had no satellite peaks, confirming that high temperatures tend to mix the superlattice sublayers. On the other hand, no films grown directly on MgO~(001) at any temperature possessed a satellite peak, suggesting that roughness caused by island growth during nucleation also plays a major role in superlattice sublayer quality. Additionally, the presence of XRD satellite peaks correlated with excellent sublayer contrast for samples also measured in STEM-EELS. Superlattice periodicity calculated from satellite peaks also agreed well with total film thickness measurements using XRR divided by the number of deposited superlattice periods.

Superlattice films grown at $\mathrm{150^{\circ}C}$, including the film shown in Fig.~\ref{EELS}(a), had additional faint RHEED diffraction spots in the [110] direction as seen in Fig.~\ref{RHEED}(a). These spots were also observed during room temperature growth of $\mathrm{Fe_2MnAl}$ and may be due to crystal twinning or the presence of nanoscale crystallites at the surface. A secondary bulk crystal phase is unlikely due to the lack of additional peaks in XRD rocking curves, and cross-sectional TEM showed no indication of surface crystal phase segregation. The extra RHEED spots vanished if the sample was annealed to $\mathrm{300^{\circ}C}$ after deposition. Spots in RHEED are often associated with bulk diffraction due to surface roughness, but RMS roughness measured in AFM was the same for samples with and without the post-growth anneal, suggesting that any changes in roughness occurred on the nanoscale.

\subsection{Magnetic Anisotropy Energy}
The effective PMA energy, $K_{eff}^{\perp}$, was experimentally quantified for each sample as the area between the out-of-plane and in-plane SQUID hysteresis loops \cite{Johnson1996}. Positive values of $K_{eff}^{\perp}$ indicated dominant PMA, while negative values indicated in-plane dominated anisotropy. $K_{eff}^{\perp}$ may be written as a sum of independent anisotropy contributions,
\begin{equation}
\label{keff}
K_{eff}^{\perp} = K_{MCA}^{\perp}+\frac{K_S}{t_{film}}-2\pi M_S^2
\end{equation}
\noindent where $K_{MCA}^{\perp}$ is out-of-plane magnetocrystalline anisotropy, $K_S$ is interface anisotropy, $t_{film}$ is the total thickness of the ferromagnetic layer, and $M_S$ is saturation magnetization. Typically, the interface anisotropy term is exploited in ultra-thin films to obtain PMA, for example in CoFeB/MgO films less than 1.5~nm thick \cite{Ikeda2010}. However, to obtain PMA from the bulk of a $t_{film}~\mathrm{=20~nm}$ thick superlattice film, two conditions must be satisfied: (i)~shape anisotropy magnitude must be minimized by reducing the saturation magnetization, and (ii)~$K_{MCA}^{\perp}$ must be maximized via tetragonal distortion and superlattice effects. 

To address condition (i), the shape anisotropy term $2\pi M_S^2$ (in~cgs~units) was reduced by increasing the aluminum content. This is possible because the $M_S$ of Heusler compounds is directly related to composition via the Slater-Pauling curve,
\begin{equation}
\label{Slater}
m=M_S/f.u.=N_V-6N_a
\end{equation}
\noindent where $m$ is the moment per molecular formula unit ($f.u.$) in units of Bohr magnetons ($\mu_B$), $N_V$ is the average number of valence electrons per $f.u.$, and $N_a$ is the number of atoms per $f.u.$ \cite{Kubler2007}. For full-Heuslers with no vacancies, (\ref{Slater}) reduces to the familiar $m = N_V-24$. However, generally, estimation of $N_V$ and $N_a$ is model-dependent, and here we choose the model with stoichiometry given by $\mathrm{Co_{2\eta}Mn_{\eta}Al_{1+x}/Fe_{2\eta}Mn_{\eta}Al_{1+x}}$, where x is aluminum excess and $\mathrm{\eta=(3-x)/3}$ is a normalization factor required to maintain four atoms per full-Heusler formula unit without vacancies, while maintaining growth fluxes as \mbox{$\Phi_{Co}=\Phi_{Fe}=2\Phi_{Mn}$}. Alternative models incorporating preferential site occupancy and vacancies ($N_a < 4$) overpredicted the reduction in $M_S$ for estimated aluminum excess based on Rutherford backscattering spectrometry calibration samples.

\begin{figure}[h]
\includegraphics[width=.9\linewidth]{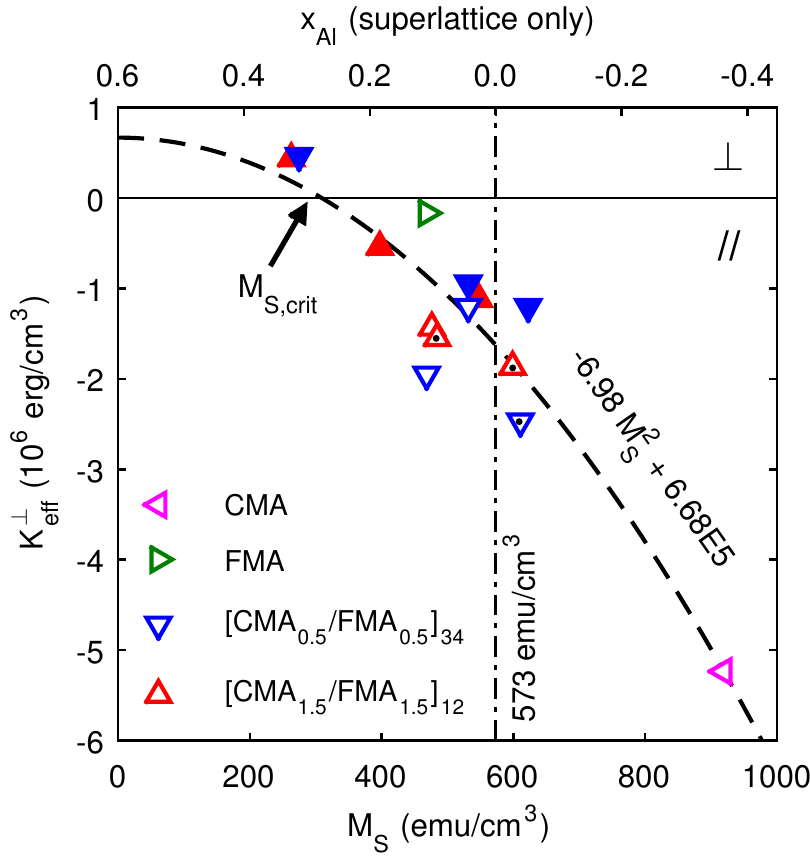}
\caption{\label{Keff}Effective perpendicular anisotropy vs. saturation magnetization at $T=5~K$. Filled markers are films grown on GaAs~(001), unfilled markers are films grown on MgO~(001), and markers with a black center dot are films grown on Cr/MgO~(001). A constrained (see text) least squares fit (black dashed line) includes all the data points shown. The upper $\mathrm{x_{Al}}$ scale indicates the calculated superlattice aluminum excess based on $M_S$ measurements and is not valid for pure CMA or FMA since they follow a different Slater-Pauling model. Out-of-plane easy axes are observed for \CMAFMA with high aluminum content grown at $\mathrm{150^{\circ}C}$ on GaAs~(001). Sample volumes were determined using XRR and photographic area measurements.}
\end{figure}

The Slater-Pauling curve for this model can be simplified to $m=\mathrm{3-5x}$, where $\mathrm{x}$ is the aluminum excess. For $\mathrm{x=0}$ and cubic lattice parameter $\mathrm{a_0=5.79}$~\AA, we expect $m=\mathrm{3.0}~\mu_B$, which gives a saturation magnetization $M_S=\mathrm{573~emu/cm^3}$. However, for 10\% aluminum excess, $M_S$ is lowered by 17\% according to the Slater-Pauling curve. This, in turn, decreases shape anisotropy magnitude by 31\%. To illustrate this point, $K_{eff}^{\perp}$ vs. $M_S$ is plotted in Fig.~\ref{Keff}. It is important to note that this figure contains data from samples grown on all three substrate types at various growth and annealing temperatures. Nevertheless, saturation magnetization is clearly an important factor in determining in-plane vs.\ out-of-plane magnetization for this system. A least-squares fit to (\ref{keff}) for the full data set is also shown. It is unclear whether the magnetocrystalline or interface anisotropy terms have some hidden dependence on $M_S$. Therefore, the linear fit term was assumed to be zero. Fits constrained to pass through the origin, as well as unconstrained fits, produced qualitatively similar results. From the fit parameters, the critical saturation magnetization was determined to be $M_{S,crit}~\mathrm{=309~emu/cm^3}$, or $m_{crit} = \mathrm{1.62}~\mu_B$. For magnetizations below $M_{S,crit}$, films were preferentially magnetized out-of-plane at temperatures below 200~K, and magnetized in-plane at temperatures above 200~K. The Curie point was above room temperature for all films measured in SQUID, so the transition observed at 200~K is a spin reorientation transition. Assuming the stoichiometry model above, $M_{S,crit}$ corresponds to an aluminum excess of $\mathrm{x>0.28}$. Therefore, PMA is observed for superlattices with greater aluminum excess than $\mathrm{Co_{1.81}Mn_{0.91}Al_{1.28}/Fe_{1.81}Mn_{0.91}Al_{1.28}}$. This aluminum excess value matches well with estimated fluxes based on RBS calibrations used for the sample set. 

To address condition (ii), $K_{MCA}^{\perp}$ was enhanced in samples with well-defined superlattice layers and tetragonal distortion $\mathrm{c/a>1}$. The two samples in Fig.~\ref{Keff} with dominant PMA ($K_{eff}^{\perp}>0$) were grown at low temperature on GaAs~(001) under conditions where superlattice sublayer interdiffusion and the degree of film relaxation was low, as described previously. A separate sample series (not shown) grown only on MgO~(001) substrates at $\mathrm{300^{\circ}C}$ did not experience a spin reorientation transition below $M_{S,crit}$, further supporting the conclusion that $K_{eff}^{\perp}$ is maximized for \CMAFMA with high aluminum excess grown on GaAs~(001) at $\mathrm{150^{\circ}C}$ with no subsequent anneal. This combines the advantages of the superlattice structure with compressive strain resulting in $\mathrm{c/a>1}$, both of which are expected to enhance $K_{MCA}^{\perp}$ \cite{Azadani2016,Burkert2004}. This factor is then allowed to dominate by increasing the aluminum content, thereby lowering the shape anisotropy contribution.

\subsection{Surface Spin Polarization}
Spin-resolved photoemission spectroscopy (SR-PES) was conducted at the Cassiop\'ee beamline at Synchrotron SOLEIL in Saint-Aubin, France. SR-PES has a probing depth of approximately 10--15~\AA\ for photoelectron kinetic energies considered here, making it an ideal technique to measure spin polarization near the surface of thin films. The samples were grown in an MBE chamber with base pressure $\mathrm{<5\times10^{-10}~Torr}$, then transferred under UHV conditions to an analysis chamber with base pressure $\mathrm{<5\times10^{-11}~Torr}$. Cobalt and iron were deposited from dual e-beam evaporators, while manganese and aluminum were deposited from standard effusion cells. Fluxes from each source were calibrated with a retractable quartz crystal microbalance in the sample position before each growth, and the microbalance tooling factors were calibrated by RBS measurements for each element. Samples were grown close to stoichiometry, resulting in films that were magnetized in-plane. Substrates and superlattice films were prepared as previously described. Iron and FMA buffer layers were grown on MgO~(001) substrates at room temperature, then annealed at $\mathrm{600^{\circ}C}$ for 20~min until \textit{in-situ} RHEED patterns became streaky, indicating a smooth surface. All superlattice samples measured in SR-PES had n~=~1.5 and were annealed at $\mathrm{300^{\circ}C}$, resulting in preferential B2 ordering as measured in XRD. However, again, bright half-order streaks were observed in RHEED, suggesting that the surface unit cell was at least partially $\mathrm{L2_1}$-like. As described earlier, these growth conditions caused sublayer intermixing. Nevertheless, the surface spin polarization was found to depend strongly on whether the superlattice was terminated with a CMA or an FMA layer, suggesting that the superlattice structure remained at least partially intact.

SR-PES measurements were performed at constant photon energy $h\nu=\mathrm{35~eV}$. Assuming that CMA and FMA possess an inner potential $V_0$ that is similar to other Heusler compounds \cite{Logan2016,Kawasaki2014}, out-of-plane photoelectron momentum $k_z$ was near the $X$ point in the bulk Brillouin zone. The analyzer was set to angle-averaging transmission mode, which integrated 52\% of the width of the surface Brillouin zone along the $\bar{X}_1$ axis (parallel to [110]), centered about the surface $\bar{\Gamma}$ point. Samples were magnetized along the Heusler~[110] direction in a 200~Oe applied field prior to each measurement, and data were collected at remanence. The Mott detector measured spin polarization in the Heusler~[110] (in-plane) and [001] (out-of-plane) directions. After measurement, samples were capped with 10~nm thick gold and loaded out of UHV for further characterization. Spin polarization was calculated as $P=A/(SR)$, where $A$ is the photoelectron scattering asymmetry in the Mott detector, $S$ is the Sherman function of the detector, and $R$ is the magnetic remanence of each sample along the Heusler~[110] direction.

\begin{figure}[ht!]
\includegraphics[width=\linewidth]{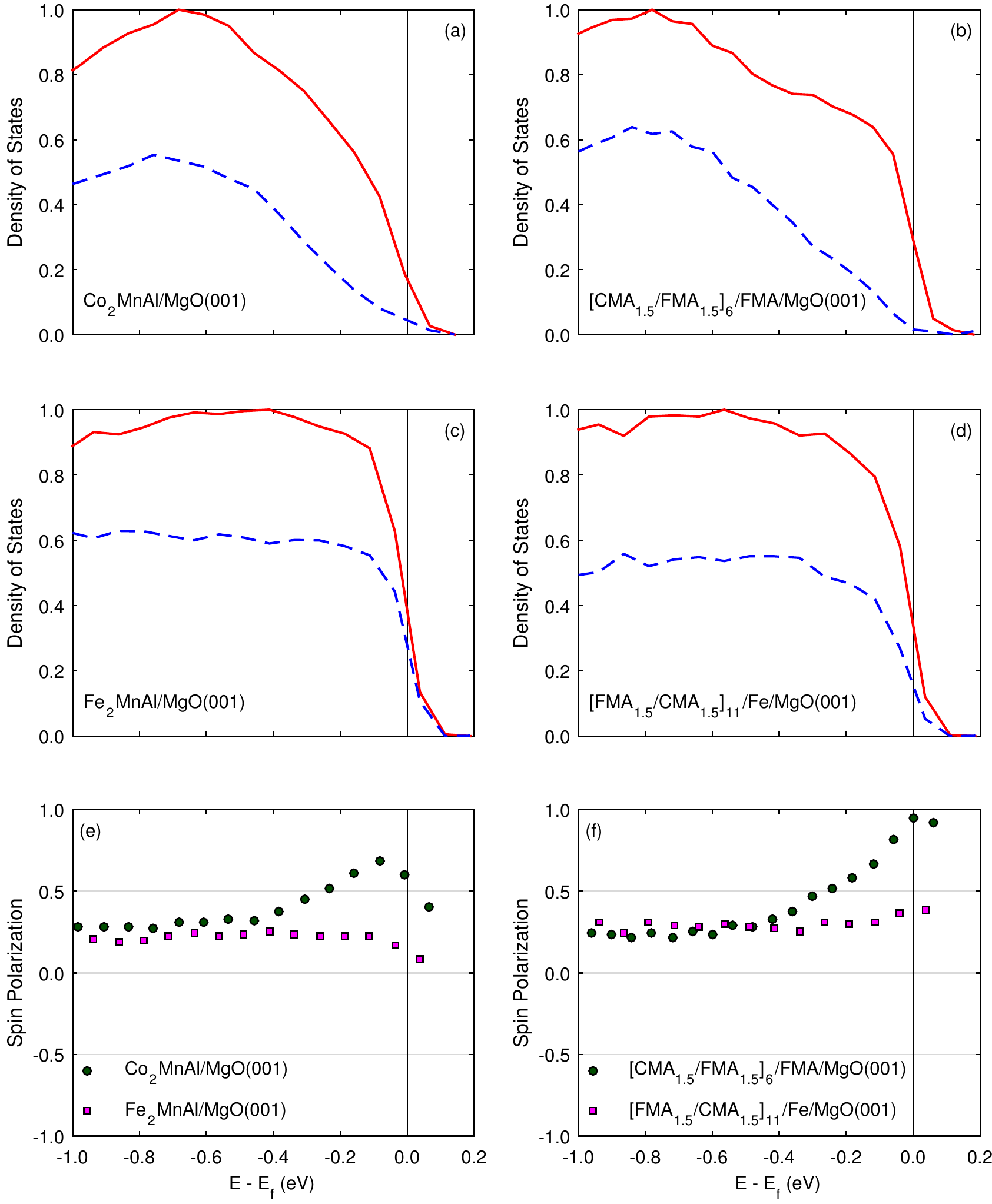}
\caption{\label{SRPES}SR-PES vs. photoelectron energy collected at $h\nu=\mathrm{35~eV}$ for four different samples, as indicated in each subplot. Normalized density of states in (a)-(d) are separated into majority (solid red line) and minority (dashed blue line) spins. The spin polarizations for pure films (a) and (c) are summarized in (e), while superlattice films (b) and (d) are summarized in (f).}
\end{figure}

From SR-PES and SQUID magnetization data, several trends emerge. First, the magnetic easy axis was found to be along [110] for CMA and \CMAFMA deposited directly on MgO (001) and GaAs (001). An easy axis along [100] was found for FMA and \CMAFMA films deposited on a 20 nm thick iron or FMA buffer layer. Furthermore, spin polarization did not depend on the substrate used, but was found to depend strongly on the surface termination layer, as shown in Fig.~\ref{SRPES}. Pure FMA films and FMA terminated \FMACMA had low spin polarization near the Fermi level, $P(E_f)=\mathrm{25\%}$, which contradicts predictions of half-metallicity for this system \cite{Fuji1995,Belkhouane2015,Azar2012,Dahmane2016}. Pure CMA had relatively high $P(E_f)=\mathrm{65\%}$, which falls short of predictions of half-metallicity for CMA \cite{Ozdogan2006,Ozdogan2011,Rai2010}, but corroborates claims of near half-metallicity with Fermi level position at the bottom of the Slater-Pauling spin gap \cite{Sakuraba2010,Li2012,Candan2013,Tung2013}.

Finally, and most importantly, CMA-terminated superlattice with $P(E_f)=\mathrm{95\%}$ shown in Fig.~\ref{SRPES}(b) and (f) had significantly higher Fermi level spin polarization than a pure CMA film. The enhancement is speculated to arise due to Fermi level tuning by the superlattice structure \cite{Azadani2016}. Additionally, given the termination dependence of the enhancement, any heterostructure interface such as superlattice/Ag for GMR devices or superlattice/MgO for tunnel junction devices should have CMA termination to maximize magnetoresistance. This conclusion is exciting because the $\mathrm{Co_2MnAl/MgO}$ interface is also expected to preserve coherent tunneling of the $\Delta_1$ Bloch band, which is a requirement for the spin filtering enhancement to tunnel magnetoresistance \cite{Miura2010,Butler2001}.

The \CMAFMA superlattice has been demonstrated to exhibit both perpendicular magnetization and near half-metallicity. However, so far these properties have been observed for samples with different growth conditions. Low saturation magnetization, low growth temperatures, and compressive substrates are required to overcome shape anisotropy and produce out-of-plane easy axes, as demonstrated for \CMAFMA with an excess of aluminum grown at $\mathrm{150^{\circ}C}$ on GaAs~(001). Near-half-metallic samples measured in SR-PES were grown close to stoichiometry on FMA/MgO~(001) substrates at $\mathrm{300^{\circ}C}$. Future work includes measuring the spin polarization of out-of-plane magnetized superlattice films with high aluminum content. Theory predicts that B2 ordering and manganese excess both preserve half-metallicity, but cobalt antisite disorder should destroy half-metallicity in CMA \cite{Ozdogan2011,Feng2015}. The effects of excess aluminum in CMA or FMA have not been reported in literature. This issue may be circumvented by growing superlattice films with $\mathrm{(Co,Fe)_2Mn_{1-x}Al_{1+x}}$ layers, which would allow for $M_S$ tuning, preserve B2 order, and minimize any potential for aluminum in the cobalt or iron sites. On the other hand, excess aluminum may be beneficial in this system as it is in $\mathrm{Co_2Mn_xSi}$, where it is argued that excess manganese prevents cobalt antisite disorder \cite{Ishikawa2009}.

\section{Conclusion}
MBE growth of single crystal epitaxial \CMAFMA superlattices on GaAs~(001), MgO~(001), and Cr/MgO~(001) substrates was successfully demonstrated. Mixed B2/$\mathrm{L2_1}$ atomic order was determined with a combination of RHEED, XRD, and HAADF-STEM. Superlattices with high sublayer structure integrity seen in STEM-EELS also possessed a weak superlattice satellite peak in XRD rocking curves. Substrate-dependent strain and tetragonal distortion was quantified by XRD RSMs, from which the relaxed cubic lattice parameter $\mathrm{a_0=5.79~\AA}$ was extracted. PMA measured in SQUID depended largely on film stoichiometry, with higher aluminum content corresponding to higher PMA. The films under the critical magnetization of $\mathrm{309~emu/cm^3}$ grown at $\mathrm{150^{\circ}C}$ substrate temperature on GaAs~(001) exhibited out-of-plane magnetization for $T<\mathrm{200~K}$. Assuming the excess aluminum is randomly substituted on cobalt, iron, and manganese atomic sublattices, this magnetization corresponds to an aluminum excess of 28\%. Synchrotron-based SR-PES measurements show the spin polarization of stoichiometric, in-plane magnetized FMA is 25\% and that of CMA is 65\% at the Fermi level near the bulk $X$ point. Superlattice \CMAFMA adopted the electronic character of the termination layer, but provided an additional improvement in spin-polarization for CMA termination, resulting in spin polarization of 95\% near the Fermi level.
\section*{Acknowledgements}
Magnetic measurements and structural characterization were supported in part by C-SPIN, one of the six centers of STARnet, a Semiconductor Research Corporation program sponsored by MARCO and DARPA. Materials growth and spin-resolved photoemission spectroscopy funding was provided by the U.S. Department of Energy (DE-SC0014388). The research reported here made use of shared facilities of the UCSB MRSEC (NSF DMR 1720256), a member of the Materials Research Facilities Network. We acknowledge valuable technical discussions with Mark Mangus and Barry Wilkens, and the rapid response of the Ion Beam Analysis of Materials facility in the LeRoy Eyring Center for Solid State Science at Arizona State University. The authors thank Fran\c coise Deschamps and Daniel Ragonnet for their assistance at the Cassiop\'ee beamline at Synchrotron SOLEIL. Finally, we thank Mihir Pendharkar, Borzoyeh Shojaei and others within the Palmstr\o m Research Group at UCSB, whose efforts made possible the preparation of GaAs substrates used in this work.
\bibliography{CFMA}

\onecolumngrid
\newpage

\begin{table}[h]
\begin{tabular}{ c | c | c | c | c | c }
\hline
\hline
Substrate & a~(\AA) & $\mathrm{\epsilon_{xx}}$~(\%) & c/a & $\mathrm{T_{g}~(^{\circ}C)}$ & $\sigma_{RMS}$~(\AA) \\
\hline
GaAs (001) & 5.653 & -2.4 & 1.02--1.06 & 150 & 8.3 \\
MgO (001) & 4.212 & 2.9 & 0.96--0.99 & 300 & 6.0 \\
Cr/MgO (001) & 2.91 & 0.5 & 1.00 & 250 & 2.4 \\
\hline
\hline
\end{tabular}
\caption{Summary of substrate lattice parameters, in-plane biaxial strain for superlattice films with $\mathrm{a_0 = 5.79}$~\AA, typical tetragonal distortion values, optimized substrate temperature during growth ($T_g$), and RMS roughness determined by AFM for 20 nm thick superlattice films.}
\label{table}
\end{table}

\newpage

\newpage

\newpage

\newpage

\newpage

\newpage

\end{document}